\documentclass[11pt, amsmath, amssymb, amsfonts]{article}

\usepackage{bm,graphicx, times}

\usepackage[english]{babel}

\usepackage{epsfig, amssymb}

\newcommand{\finpr}{\hfill $\square$ \vspace{2mm}}

\def\be{\begin{eqnarray}}
\def\ee{\end{eqnarray}}
\def\bee{\begin{eqnarray*}}
\def\eee{\end{eqnarray*}}
\newtheorem{thm}{Theorem}

\newtheorem{lem}{Lemma}

\begin{document}
\title{\bf Classical simulation of quantum computation, the Gottesman-Knill theorem, and slightly beyond}

\author{Maarten Van den Nest\footnote{Max-Planck-Institut f\"ur Quantenoptik, Hans-Kopfermann-Str. 1, D-85748 Garching, Germany.}
}

\maketitle

\begin{abstract}
We study classical simulation of quantum computation, taking the Gottesman-Knill theorem as a starting point. We show how each Clifford circuit can be reduced to an equivalent, manifestly simulatable circuit (normal form). This provides a simple proof of the Gottesman-Knill theorem without resorting to stabilizer techniques. The normal form highlights why Clifford circuits have such limited computational power in spite of their high entangling power. At the same time, the normal form shows how the classical simulation of Clifford circuits fits into the standard way of embedding classical computation into the quantum circuit model. This leads to simple extensions of Clifford circuits which are classically simulatable. These circuits can be efficiently simulated by classical sampling (``weak simulation'') even though the problem of exactly computing the outcomes of measurements for these circuits (``strong simulation'') is proved to be $\#\mathbf{P}$-complete---thus showing that there is a separation between weak and strong classical simulation of quantum computation.
\end{abstract}

\section{Introduction}

The Gottesman-Knill theorem \cite{Go98} is a well-known result in quantum information theory which states that a certain class of non-trivial quantum circuits, called ``Clifford circuits'', can be simulated efficiently on a classical computer, and can hence not provide any speed-up w.r.t. classical computation. A Clifford circuit is any quantum circuit which is composed of Hadamard, PHASE and CNOT gates. The Gottesman-Knill theorem asserts that each (uniform family of) Clifford circuit(s), when acting on the computational basis state $|\mathbf{0}\rangle\equiv |0\rangle^N$, and when followed by a computational basis measurement, can be simulated efficiently on a classical computer.

While in fact not so hard to prove, this result exhibits some rather remarkable and sometimes puzzling features, not all of which are fully understood. For example, even though they are efficiently classically simulatable, Clifford circuits can generate a high degree of entanglement \cite{He06}; the highly entangled cluster states \cite{Br01} can e.g. be generated by Clifford circuits. This very feature raises doubts about the often-recited mantra that ``entanglement is responsible for the quantum computational speed-up''. In particular, it highlights that, while the presence of (certain types of) entanglement in a quantum computation (QC) is provably necessary to disallow efficient classical simulation (see e.g. \cite{Jo02, Vi03, Ma05, Sh06, Yo06, Jo06, Va06}), it is certainly not sufficient.

Further, it is known  that Clifford circuits can efficiently be simulated classically by a rather restricted classical computer, namely a circuit model computer which only uses NOT and CNOT gates \cite{Aa04}. In other words, it is not necessary to invoke the full power of classical computers to efficiently simulate arbitrary Clifford circuits. Hence, Clifford circuits are most likely not even universal for \emph{classical} computation. In complexity theoretic terms, the power of classical computation with NOT and CNOT gates---and hence of Clifford circuits---is captured by the complexity class $\oplus\mathbf{L}$ (``parity-L'') \cite{Pa82, Da90}. This class is known to be contained in $\mathbf{P}$ but not expected to be equal to it (although this is an unproven conjecture). Nevertheless, supplementing Clifford operations with essentially any non-Clifford gate immediately yields the full QC model \cite{Ne01, So00}. This yields an immediate ``jump'' in computational power from $\oplus \mathbf{L}$ to $\mathbf{BQP}$ rather than a ``smooth'' transition $\oplus\mathbf{L}\to\mathbf{BPP}\to\mathbf{BQP}$. In particular, this property makes it hard to extend Clifford circuits to a class of efficiently simulatable quantum circuits which has the same computational power as full classical computation.

It is the aim of this note to obtain some insight in the above list of features. First we will show that each Clifford circuit  ${\cal C}$ can be (efficiently) reduced to an equivalent circuit ${\cal C}'$ which yields the same output. The ``normal form'' ${\cal C}'$ is also a Clifford circuit but has a very simple structure, as displayed in Fig. \ref{Fig:Clifford_circuit}; this normal form is based on an earlier result \cite{De03}, for which we also provide a simple proof. We will see that ${\cal C}'$ is manifestly efficiently classically simulatable. We argue that the normal form also sheds some light on why the high degree of entanglement and interference generated by Clifford operations does not result in any (exponential) quantum computational speed-up. Finally, the normal form shows how the simulation of Clifford operations fits within the standard embedding of probabilistic classical computation into the quantum circuit model. The latter is related to the notion of HT circuits, as discussed in section \ref{sect_BPP_BQP}.

Along the way, we will make some general remarks regarding classical simulation of QC. One of them regards the different possible definitions of the notion of ``classical simulation''. When a QC is to be simulated classically, the aim may be to either (i) \emph{compute} the probabilities of the output measurement efficiently classically with high accuracy (``strong simulation'') or (ii) \emph{sample} from this distribution efficiently using a classical computer (``weak simulation''). Both variants constitute valid classical simulation techniques; however, a priori it is not clear whether there is a clear separation between these two notions. The vast majority of all works (see e.g. \cite{Go98, Ma05, Sh06, Jo06, Va06, Va01, Di04, Br08, Jo08, Ah06, Yo07, Br07, Bra07}) regarding classical simulation of QC considers quantum circuits which can be simulated efficiently classically in the strong sense, whereas not much is known about weak simulation.

We will find that the simulatable quantum circuits considered  in the present paper are \emph{not} amenable to strong simulation---in fact, it is easy to show that strong simulation of these circuits constitutes a $\#\mathbf{P}$-complete problem. Nevertheless, weak classical simulation of the same circuits is efficiently possible by means of a very simple sampling technique. This shows that strong and weak simulation of QC are fundamentally different notions. Moreover, these examples highlight that any serious attempt to understand the relationship between classical and quantum computation should not rely on the notion of strong simulation.

\section{Classical simulation of quantum computation}

Here we state more precisely what we mean by ``efficient classical simulation of quantum computation''. We essentially follow the definitions from \cite{Jo02, Jo08}.

Consider a uniform family of quantum circuits\footnote{In the following, when we refer to a ``quantum circuit'' we will always mean a uniform family of quantum circuits.} ${\cal U}\equiv {\cal U}_N$ acting on the $N$-qubit input state $|\mathbf{0}\rangle\equiv |0\rangle^{\otimes N}$, and followed by a measurement of, say, the first qubit in the computational basis. Then this quantum computation yields as an outcome a bit $\alpha\in\{0, 1\}$. The probability that the outcome $\alpha$ occurs is given by $\pi(\alpha)=\langle\mathbf{0}|{\cal U}^{\dagger}[|\alpha\rangle\langle\alpha|\otimes I]{\cal U}|\mathbf{0}\rangle$. We say that the above quantum computation can be efficiently simulated classically in the \emph{strong} sense if it is possible to evaluate $\pi(0)$ up to $M$ digits in poly$(N, M)$ time on a classical computer. Furthermore, we say that the quantum computation can be efficiently simulated classically in the \emph{weak} sense if it is possible to sample once from the probability distribution $\{\pi(\alpha)\}$ in poly$(N)$ time on a classical computer\footnote{More precisely, one should require to sample from a probability distribution which is not necessarily exactly equal to $\{\pi(\alpha)\}$, but sufficiently close to it (see e.g. \cite{Jo02}). In the examples in this paper, however, it will always be possibly to perform an exact sampling, such that we omit such accuracy issues for simplicity.}.

We further point out that to date the vast majority of results on simulation of quantum computation regard \emph{strong} classical simulation. This regards e.g. the Gottesman-Knill theorem, the simulation of ``matchgates'' \cite{Va01, Di04, Br08, Jo08}, the simulation of the quantum Fourier transform \cite{Ah06, Yo07, Br07}, simulation results involving tensor contracting techniques \cite{Ma05, Sh06, Jo06, Va06}, etc. However, in order to study the difference between classical and quantum computation, the notion of weak classical simulation is much more appropriate. In fact, the strong simulation of general quantum circuits can easily be shown to be a $\#\mathbf{P}$-hard problem (this can  e.g. been showed by considering the the family of circuits $U_f$ as in section \ref{sect_BPP_BQP}). As $\mathbf{BQP}$ is believed to be much smaller than $\#\mathbf{P}$, this indicates that strong classical simulation inevitably seems too strong a requirement to study the relation between classical and quantum computation. However, not much is known about the difference between strong and weak simulation of quantum computation. In section \ref{sect_BPP_BQP} we will give an example of a class of quantum circuits which is efficiently simulatable in the weak sense but which is intractable in the strong sense (unless $\mathbf{P}$ is equal to $\#\mathbf{P}$).

\section{The Gottesman-Knill theorem}\label{sect_GK}

In this section we recall the Gottesman-Knill theorem and  the relation between Clifford operations and the complexity class $\oplus\mathbf{L}$. (A variant of) the Gottesman-Knill theorem can be formulated as follows \cite{Go98}:

\begin{thm}[Gottesman-Knill] Every (uniform family of) Clifford circuit(s), when applied to the input state $|\mathbf{0}\rangle\equiv|0\rangle^{\otimes N}$ and when followed by a $Z$ measurement of the first qubit, can be efficiently simulated classically in the strong sense.
\end{thm}
The standard technique to prove this result involves a connection between Clifford operations and groups of commuting Pauli operations called ``stabilizer groups''. We omit this proof here and refer to e.g. \cite{Go98, Ni00}. Below we will provide an alternative proof of the Gottesman-Knill theorem which does not use stabilizer techniques.

It turns out that a rather  restricted classical computer suffices  to efficiently simulate arbitrary Clifford circuits. Indeed,  every Clifford circuit can be efficiently simulated by a classical circuit model computer which uses NOT and CNOT gates only, applied to the input string $\mathbf{0}=(0, \dots, 0)$ \cite{Aa04}. More precisely, in \cite{Aa04} the problem {\sc Gottesman-Knill} was defined as follows: the input is an $n$-qubit Clifford circuit ${\cal C}$; the problem is to decide whether the first qubit will be in the state $|1\rangle$ with certainty after ${\cal C}$ has been applied to the input $|0\rangle^n$. It was showed in \cite{Aa04} that this problem can be mapped, under a logarithmic-space reduction, to a problem of simulating a classical poly-size CNOT-NOT circuit acting on the all-zeroes input state. The complexity class of problems that are log-space reducible to the simulation of CNOT-NOT circuits, is called $\oplus\mathbf{L}$ (``Parity-$\mathbf{L}$'') \footnote{Obviously, $\oplus\mathbf{L}\subseteq \mathbf{P}$; the inclusion is believed to be strict, but a proof of this conjecture has not yet been found.} \cite{Pa82, Da90}. Hence, it was showed in \cite{Aa04} that {\sc Gottesman-Knill} is in $\oplus\mathbf{L}$. As CNOT and NOT gates are Clifford operations, it is clear that every $\oplus\mathbf{L}$ problem can be reduced to {\sc Gottesman-Knill}. In conclusion, the problem {\sc Gottesman-Knill} is \emph{$\oplus\mathbf{L}$-complete} \cite{Aa04}.

The class of problems in $\oplus \mathbf{L}$ are centered around linear algebra over the finite field $\mathbf{Z}_2$. It has been shown that the problems of solving linear equations over $\mathbf{Z}_2$, finding the inverse of a nonsingular $\mathbf{Z}_2$-matrix, multiplying matrices over $\mathbf{Z}_2$ etc., are $\oplus\mathbf{L}$  problems \cite{Da90}.

\section{Embedding classical in quantum computation}\label{sect_BPP_BQP}

Here we  briefly review how classical (probabilistic) computation can be regarded as being a ``part'' of quantum computation. All material in this section is standard (see e.g. \cite{Ni00}), except for an observation about the difference between strong and weak simulation of quantum computation, which will be made at the end of this section.

\begin{figure}
\hspace{3.5cm} {\includegraphics[width=7cm]{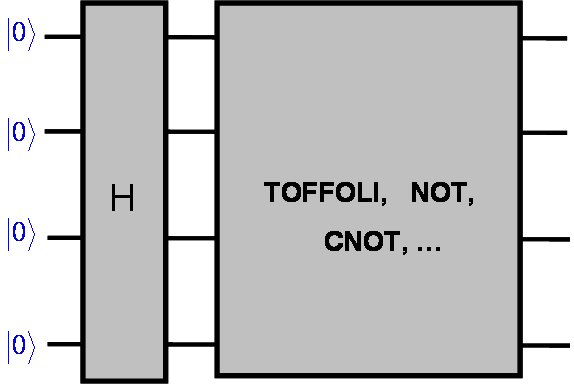}}
\caption[]{\label{Fig:Toffoli_circuit}
The class of quantum circuits with the above structure are here called HT circuits (first a round of {\sc Hadamard} gates is applied to a subset of the qubits, followed by a round of ``classical gates'', i.e. {\sc Toffoli}, {\sc not}, etc.). HT circuits followed by a computational basis measurement have the same computational power as probabilistic classical computation: each probabilistic classical computation can be simulated efficiently by an HT circuit, and each HT circuit can be weakly simulated efficiently on a classical computer. However, strong classical simulation of HT circuits is a $\#\mathbf{P}$-complete problem.}
\end{figure}

Probabilistic classical computation is classical computation (e.g.  considered in the circuit model) supplemented with the possibility of deciding, in each step of the computation, which gate to apply based on the random outcome of a coin toss. It is well understood how probabilistic classical computation can be embedded in the full pure-state circuit model of quantum computation. The subclass of quantum circuits that corresponds to  probabilistic classical computation is the following: (ROUND 1) apply Hadamard gates to an arbitrary subset of qubits; (ROUND 2) apply a (uniform family of) circuit(s) consisting of classical\footnote{We call any unitary gate $U$ a ``classical gate'' if it maps computational basis states to computational basis states.} gates only (e.g. NOT, CNOT, Toffoli gates).
A quantum circuit with the above structure will be called an HT circuit (short for Hadamard-Toffoli); see also Fig \ref{Fig:Toffoli_circuit}. It can easily be shown that every probabilistic classical computation can be accounted for by a suitable HT circuit acting on the all-zeroes input state and followed by a single-qubit $Z$ measurement (say, on the first qubit). Conversely, it is also straightforward to see that each HT circuit (with input and measurement as above) can be simulated efficiently---in the weak sense---by a probabilistic classical computer: if the quantum register has $N$ qubits, if $m$ denotes the number of Hadamard gates applied in ROUND 1, and if the classical gates in ROUND 2 compute an efficiently computable (invertible)  function $f:\{0, 1\}^N\to\{0, 1\}^N$, then the state of the system after the first two rounds has the form $\sum_{x\in\{0, 1\}^m} |f(x, 0, \dots, 0)\rangle,$ where there are $N-m$ zeroes in the argument of $f$. We will use the abbreviation $F(x)\equiv f(x,0, \dots, 0)$. 
Performing a $\{|0\rangle, |1\rangle\}$ measurement on the first qubit yields as an outcome a bit $\alpha\in\{0, 1\}$. The probability that  $\alpha$ occurs is equal to \be\label{pi_alpha} \pi(\mathbf{\alpha}) = \frac{|\{x\in \{0, 1\}^m: F(x)_{1}=\alpha\}|}{2^m},\ee where $F(x)_1$ denotes the first bit of $F(x)$. Sampling from this probability distribution can easily be done classically: simply generate an $m$-bit string $x$ uniformly at random, then compute $F(x)$ and finally set $\alpha:=F(x)_1$; then $\alpha$ is generated with probability $\pi(\alpha)$ as desired. This shows that weak classical simulation of HT quantum circuits is efficiently possible. We can therefore conclude that the computational power of HT quantum circuits is equivalent to probabilistic classical computation.

We remark that \emph{strong} classical simulation of HT circuits is much harder---in fact, we point out that such strong simulation is a $\#\mathbf{P}$-complete problem. The complexity class $\#\mathbf{P}$ is concerned with \emph{counting problems}: given an efficiently computable Boolean function $f\in $ P, the problem of determining the integer $\#f:=|\{x: f(x)=0\}|$ defines the complexity class $\#\mathbf{P}$. It is easy to see that the problem of computing the probabilities (\ref{pi_alpha}) is at least as hard as computing  $\#f$ for arbitrary functions $f\in \mathbf{P}$. To show this, consider an arbitrary Boolean function $f:\{0, 1\}^m\to\{0,1\}$ which is in $\mathbf{P}$. Then there exists a poly-sized HT quantum circuit ${\cal U}$ such that ${\cal U}|\mathbf{0}\rangle \propto \sum_x |x\rangle|f(x)\rangle.$ Performing a measurement in the computational basis on the last qubit yields a single bit with probability distribution $\pi(0) = \#f/2^{m} = 1-\pi(1).$ Computing these probabilities with perfect accuracy is thus equivalent to computing $\#f$. This shows that strong simulation of HT circuits is at least as hard as any problem in $\#\mathbf{P}$ an is thus $\#\mathbf{P}$-hard. Moreover, it can easily be shown that, for arbitrary HT circuits and for arbitrary computational basis measurements, the problem of computing the probabilities (\ref{pi_alpha}) is always in $\#\mathbf{P}$. This shows that the strong simulation of HT circuits is a $\#\mathbf{P}$-complete problem. 
In conclusion, \emph{efficient strong classical simulation of HT quantum circuits is intractable (unless $\mathbf{P}$ is equal to $\#\mathbf{P}$), while weak simulation of these circuits is efficiently possible} (even trivially so).

\section{Clifford circuits as HT circuits}

In this section we show that, modulo  some ``redundancies''  (to be explained below), each Clifford circuit can be reduced to an HT circuit, which is manifestly classically simulatable. To this aim, we provide an alternative proof of the Gottesman-Knill theorem which will allow to make the connection to HT circuits. The proof will not be centered around stabilizer groups (and hence does not require any knowledge of the stabilizer formalism) but will simply consist of  tracking the coefficients $\psi_x$ of the state $|\psi\rangle=\sum_x\psi_x|x\rangle$ of the quantum register throughout the application of an arbitrary Clifford circuit. To achieve this, it will be useful to know what the most general form is which the coefficients $\psi_x$ may assume. The latter issue has in fact already been fully investigated and understood in \cite{De03}, where the following result was proved: every $N$-qubit  state $|\psi\rangle={\cal C}|0\rangle^N$, where ${\cal C}$ represents a Clifford circuit, is given by an expression of the form \be\label{expansion} |\psi\rangle \propto \sum_{x\in A} i^{l(x)} (-1)^{q(x)} |x\rangle, \ee where we have the following notations. First, $A\subseteq \mathbf{Z}_2^N$ is an affine subspace, i.e. a subset  of the form $A = \{Ru+t| u\in\mathbf{Z}_2^m\}$, for some fixed (nonsingular) $N\times m$ matrix $R$ and vector $t\in \mathbf{Z}_2^N$.  Second, $l(x)$ is a linear function on $\mathbf{Z}_2^N$, i.e., $l$ maps $x$  to $l(x)=d^Tx$ for some $d\in\mathbf{Z}_2^N$. The exponent $l(x)$ of $i$ is computed modulo two. 
Finally, $q(x)$ is a quadratic function on $\mathbf{Z}_2^N$. That is, $q$ maps  $N$-bit strings $x=(x_1, \dots, x_N)$ to $q(x)=\sum c_{ij} x_ix_j + c_ix_i$, for some (fixed) $c_{ij}, c_i\in\{0, 1\}$.
Moreover, $q$, $l$ and $A$ can be efficiently computed. Conversely, every state of the form (\ref{expansion}) is a stabilizer state.

The proof  that an arbitrary Clifford circuit composed always maps $|\mathbf{0}\rangle$ to a state of the form (\ref{expansion}) was first derived in \cite{De03} and is based on mappings between Clifford operations, the stabilizer formalism and arithmetic over $\mathbf{Z}_2$. Here we will give a simple alternative proof of this theorem.

\textbf{Proof of Eq. (\ref{expansion}):} The proof   can be obtained by induction on  the number of gates in the circuit ${\cal C}$. Evidently, (\ref{expansion}) is true when ${\cal C}$ is the identity. Now suppose that the result is true for every Clifford circuit consisting of $K$ gates. We then have to prove that the result still holds for circuits with one additional gate. Consider an arbitrary circuit ${\cal C}$ of size $K+1$, and write ${\cal C}$ as a product of a single Clifford gate $U\in\{H,\ P,\ \mbox{CNOT}\}$ and a circuit ${\cal C}'$ of size $K$: ${\cal C} = U {\cal C}'$. By the induction step, we may assume that there exist $A$, $l$ and $q$ such that ${\cal C}'|\mathbf{0}\rangle$ is given by (\ref{expansion}). It is then simply a matter of verifying that the general form (\ref{expansion}) is kept when a gate from the set $\{H,\ P,\ \mbox{CNOT}\}$ is applied. This can be shown using straightforward arithmetic, and for completeness these calculations are given in the appendix.
\hfill $\square$

The above proof of expression (\ref{expansion}) leads to a  simple algorithm to compute $A$, $l$ and $q$ if a circuit ${\cal C}$ is given: first, one initializes $A$, $l$ and $q$ to their trivial values, corresponding to the state $|\mathbf{0}\rangle$. Then one sequentially updates $A$, $l$ and $q$ corresponding to the first, second, etc.  gate in ${\cal C}$. The final values for $A$, $l$ and $q$ then correspond to the state ${\cal C}|\mathbf{0}\rangle$.

It is now easy to see that each Clifford computation can be simulated efficiently. If  a circuit ${\cal C}$ is given, the first step is to compute the triple $(A, l, q)$ parameterizing ${\cal C}|\mathbf{0}\rangle$. The goal is then to efficiently simulate $\{|0\rangle, |1\rangle\}$ measurements on this state. To do so,  one argues at follows. First, the outcome probabilities of a measurement in the computational basis are \emph{independent} of the phases $i^{l(x)}(-1)^{q(x)}$.  That is, the specific values of $l$ and $q$ are in this context completely redundant, such that we may set these quantities to their trivial values $l\equiv 0, q\equiv 0$.
 This leaves us with the state $\sum_{x\in A}|x\rangle$. Let the affine space $A$ be given by $A = \{Ru + t: u\in\mathbf{Z}_2^m\}$, for some invertible $N\times m$ matrix $R$ and $t\in\mathbf{Z}_2^N$. Further, suppose that the first qubit is measured, yielding as an outcome a bit $\alpha$ which occurs with probability \be \pi({\alpha}) = \frac{|\{x\in A: x_{1} = \alpha\}|}{|A|}.\ee It is trivial to sample from this probability distribution: simply generate a uniformly random $m$-bit string $u$, compute $Ru+t$ and set $\alpha$ to be the first coefficient of $Ru+t$. This shows that each Clifford circuit can efficiently be simulated in the weak sense.

However, one can do more than this, as the probabilities $\pi(\alpha)$ can be \emph{computed} efficiently as well. Indeed, each of these probabilities has the form $2^{-\gamma}$ for some integer $\gamma$ which can be obtained by solving a system of linear equations over $\mathbf{Z}_2$. Hence, strong simulation is possible as well, thus recovering the Gottesman-Knill theorem.

\section{Normal form}

\begin{figure}
\hspace{1cm} {\includegraphics[width=12cm]{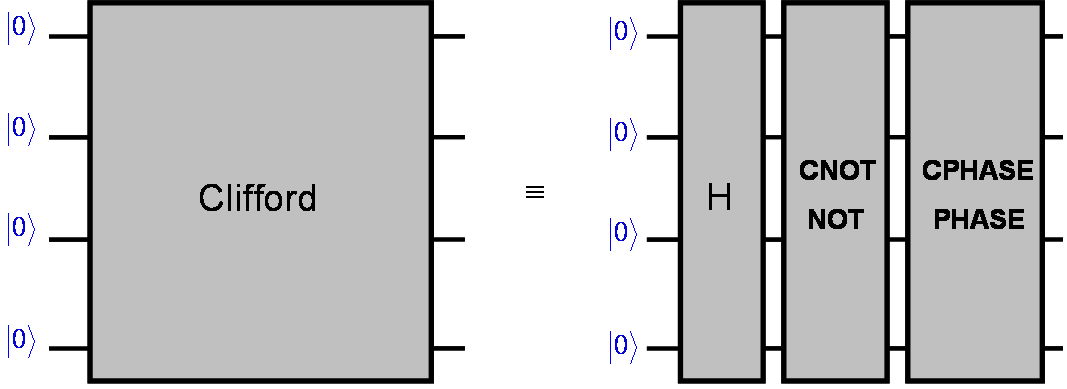}}
\caption[]{\label{Fig:Clifford_circuit} For each Clifford circuit ${\cal C}$ acting on the computational basis state $|\mathbf{0}\rangle$ (left), there exists a normal form ${\cal C}'$ (right) of the above structure such that ${\cal C}|\mathbf{0}\rangle = {\cal C}'|\mathbf{0}\rangle$, i.e., these two circuits yield the same output state. The last round, consisting of diagonal gates, is  undetected by a computational basis measurement and can therefore be completely disregarded. Moreover,  the first two rounds in ${\cal C}'$ constitute a (restricted) HT circuit and hence are trivially classically simulatable.}
\end{figure}

Next we note that the expression (\ref{expansion}) provides an alternative way---different than the circuit ${\cal C}$---to prepare the output state ${\cal C}|\mathbf{0}\rangle$. If the affine space $A$ is given by $A = \{Ru + t: u\in\mathbf{Z}_2^m\}$, then this  state can be prepared from $|\mathbf{0}\rangle$ as follows.  First, apply $m$ Hadamard gates such as to yield $\sum_u |u\rangle|0\rangle^{N-m}$. Second, apply appropriate CNOT gates such as to yield $\sum_u |Ru\rangle$. Third, apply the NOT operation $X^{t_1}\otimes\dots X^{t_N}$, yielding $\sum_u |Ru + t\rangle$. Fourth, apply appropriate PHASE and CPHASE gates such as to yield ${\cal C}|\mathbf{0}\rangle$. The overall structure of this preparation method is illustrated in Fig. \ref{Fig:Clifford_circuit}. We have now arrived the following result.

\begin{thm}\label{thm_clifford}
Let ${\cal C}$ be an arbitrary poly-sized Clifford circuit. Then there exists a poly-sized  Clifford circuit ${\cal C}'$ satisfying ${\cal C}|\mathbf{0}\rangle = {\cal C}'|\mathbf{0}\rangle$ such that ${\cal C}'$ can be decomposed into three ``rounds'': (ROUND 1) apply Hadamard gates to an arbitrary subset of qubits; (ROUND 2) apply a poly-sized circuit of NOTs and CNOTs; (ROUND 3) apply a poly-size circuit of PHASEs and CPHASEs. The circuit ${\cal C}'$ can be efficiently determined.
\end{thm}
It is important to remark that  it is generally not true that the circuits ${\cal C}$ and ${\cal C}'$ are equal as $2^N\times 2^N$ matrices\footnote{A counterexample is e.g. given by the single-qubit circuit ${\cal C}$ = HPH. By exhaustive enumeration of all possibilities, it can easily be shown that no single-qubit circuit ${\cal C}'$ of the above structure satisfies ${\cal C}={\cal C}'$.}. These circuits merely have the same effect on the input state $|\mathbf{0}\rangle$, i.e., \emph{they generate the same output state}. Note that it is the output state $|\psi_{\mbox\scriptsize{out}}\rangle={\cal C}|\mathbf{0}\rangle$ of the computation which is relevant for our purposes---i.e. a single column of the $2^N\times 2^N$ matrix ${\cal C}$---and not the entire circuit.  We further emphasize that the choice  of a standard input state $|\mathbf{0}\rangle$ as opposed to arbitrary products of $X$, $Y$, $Z$ eigenstates as inputs, does not entail any loss of generality, as such alternative input states can always be rotated into the state $|\mathbf{0}\rangle$ using a (local) Clifford operation.

The Clifford normal form highlights the computational ``weakness'' of Clifford circuits. First, the original Clifford circuit, which may contain many Hadamard gates located at different places in the circuit, causing subsequent rounds of constructive and destructive interference, is (efficiently) mapped to the normal form which does not display any interference at all. Indeed, up to the irrelevant last round of diagonal gates,  each Clifford circuit is reduced to a circuit of NOT and CNOT gates applied to a superposition of computational basis states $|+\rangle^K|0\rangle^L$---in other words, nothing but a (very simple instance of an) HT circuit, which are trivially classically simulatable. Also the connection between Clifford circuits and the complexity class $\oplus \mathbf{L}$ is in this way highlighted, given the relation between the class $\oplus\mathbf{L}$ and circuits consisting of NOT and CNOT operations.

With a little extra work, theorem \ref{thm_clifford} may be used to arrive at a normal form for Clifford circuits ${\cal C}$ which does regard the entire  $2^N\times 2^N$ matrix. The normal form is in fact highly similar to theorem \ref{thm_clifford}. The significance of this normal form is again that it contains only a single round of basis-changing operations (i.e. Hadamards). In fact, we will see that every Clifford operation can be written as a a tensor product of Hadamard operations ${\cal H} = H^S\otimes I$ acting nontrivially on a subset $S$ of the qubits, multiplied on the left and the right with  \emph{basis-preserving} Clifford circuits $M_1$ and $M_2$, i.e. circuits composed of {\sc cnot}, {\sc phase} and {\sc cphase} gates. Similar to (\ref{expansion}), (a variant of) theorem \ref{normal-Cliff2} was proved in \cite{De03} using mappings between the stabilizer formalism, Clifford operations and $\mathbb{Z}_2$-arithmetic; here we provide a direct proof.

\begin{thm}\label{normal-Cliff2}
Let ${\cal C}$ be an arbitrary $n$-qubit Clifford operation. Then there exist: (a)  poly-size circuits $M_1$ and $M_2$ composed of {\sc cnot}, {\sc phase} and {\sc cphase} gates and (b) a tensor product of {\sc Hadamard} gates and identities ${\cal H} = H^S\otimes I$ acting nontrivially on a subset $S$ of the qubits, such that ${\cal C} \propto M_2 {\cal H} M_1$. Moreover, $M_1$, $M_2$ and ${\cal H}$ can be determined efficiently.
\end{thm}
{\it Proof: } Let $a=(a_1, \dots, a_n)$ denote  an arbitrary
$n$-bit string. For every $i=1, \dots, n$, define $\sigma_i:=
{\cal C}X_i{\cal C}^{\dagger}$ where $X_i$ denotes the Pauli matrix $X$ acting on qubit $i$. Since ${\cal C}$ is a Clifford operation, each $\sigma_i$ is a Pauli operator\footnote{A Pauli operator has the form $P=P_1\otimes\dots\otimes P_n$ where each $P_i$ is either the identity or one of the Pauli matrices $X$, $Y$ or $Z$.}, possibly
with an overall minus sign. Denoting $X(a) = X^{a_1}\otimes\dots\otimes X^{a_n}$ and $\sigma(a): = \prod_i
\sigma_i^{a_i}$, we thus have $\sigma(a):= {\cal C}X(a){\cal C}^{\dagger}$. Since $|a\rangle = X(a)|0\rangle^n$, we have ${\cal C}|a\rangle = {\cal C}X(a)|0\rangle^n = \sigma(a)
{\cal C}|0\rangle^n$.
We can now apply theorem \ref{thm_clifford} to the state ${\cal C}|0\rangle^n$: there exists a poly-size Clifford circuit $M$ consisting of {\sc phase}, {\sc cphase} and {\sc cnot} gates, and an operation ${\cal H} = H^S \otimes I$ such that ${\cal C}|0\rangle^n = \gamma M {\cal H}|0\rangle^n,$
where $\gamma$ is  some overall phase factor. Now define $\tau_i :
= {\cal H} M^{\dagger}\sigma_i M {\cal H}$ for every $i$; also
$\tau_i$ is a Pauli operator, possibly supplemented with an
overall minus sign. Denoting $\tau(a): = \prod_i \tau_i^{a_i}$, we
thus have $\tau(a):= {\cal H} M^{\dagger}\sigma(a) M {\cal H}$.
This leads to \be\label{proof_normal3} {\cal C}|a\rangle = \gamma \sigma(a) M {\cal H}|0\rangle^n= \gamma M{\cal
H} \tau(a)|0\rangle^n.\ee Note that $\gamma$ does not depend on
$a$.

We now study the general  form of the unitary mapping
$|a\rangle\to \tau(a)|0\rangle^n$. Since each $\tau_i$ is an
element of the Pauli group, there exist $n$-bit strings $u=(u_1,
\dots, u_n), v=(v_1, \dots, v_n)$, $R^{i} = (R_1^{i}, \dots,
R_n^{i})$ and $T^{i} = (T_1^{i}, \dots, T_n^{i})$ such that
$\tau_i = (-1)^{u_i} i^{v_i} X(R^{i}) Z(T^{i})$ and hence
\be\label{proof_normal4} \tau(a) = \prod_i (-1)^{a_iu_i}
i^{a_iv_i} X(a_iR^{i}) Z(a_iT^{i}).\ee For every two $x,
z\in\mathbb{Z}_2^n$, we have $X(x)Z(z) = (-1)^{x^Tz} Z(z)X(x)$.
Hence, we may reshuffle the factors in the product
(\ref{proof_normal4}) at the cost of an overall phase factor of
the form $(-1)^{q(a)}$, for some quadratic function
$q:\mathbb{Z}_2^n\to \mathbb{Z}$. In particular, we may write \be
\tau(a) = (-1)^{q(a)} \prod_i i^{a_iv_i} \prod_i X(a_iR^{i})
\prod_iZ(a_iT^{i}).\ee Letting $R$ ($T$) denote the $n\times n$
matrix with columns $R^{i}$ ($T^{i}$), we then have \be \tau(a) =
(-1)^{q(a)} \prod_i i^{a_iv_i} X(Ra) Z(Ta).\ee The action of
$\tau(a)$ on $|0\rangle^n$ then has the following form: \be
\tau(a)|0\rangle^n &=& (-1)^{q(a)} \prod_i i^{a_iv_i} X(Ra)
Z(Ta)|0\rangle^n= (-1)^{q(a)} \prod_i i^{a_iv_i} |Ra\rangle,\ee
where we have used that $Z(Ta)|0\rangle^n =|0\rangle^n$ and that
$X(Ra) |0\rangle^n = |Ra\rangle$. It then easily follows that
there exists a  Clifford circuit $M'$ (independent of $a$) composed of {\sc cnot}, {\sc phase} and {\sc cphase} gates
such that $\tau(a)|0\rangle^n= M'|a\rangle$. Together with
(\ref{proof_normal3}), this completes the proof. \finpr

\section{Slightly beyond Gottesman-Knill}

The Clifford normal form also makes it easy to extend the Gottesman-Knill theorem. Next we provide a class of quantum circuits which are efficiently simulatable classically (in the weak sense), and which are extensions of Gottesman-Knill in the sense that these circuits can efficiently generate any output state which can be efficiently generated by arbitrary Clifford circuits. Moreover, contrary to Clifford circuits, the extensions encompass full probabilistic classical computation.

We consider circuits of the following structure: first, apply an arbitrary local unitary operation; afterwards, apply an arbitrary (uniform family) of quantum circuit(s) consisting of Toffoli and diagonal gates only. We show that such circuits, when applied to the input $|\mathbf{0}\rangle$ and when followed by a computational basis measurement, can be efficiently simulated classically in the weak sense. After the first round, the state of the quantum register is a complete product state $|\chi\rangle=|\chi_1\rangle\otimes\dots\otimes|\chi_N\rangle$. Denoting $|\chi_i\rangle = a_i|0\rangle +b_i|1\rangle$, the coefficients $\chi_x$ of $|\chi\rangle$ in the computational basis (where $x=(x_1, \dots, x_N)$ is an $N$-bit string) are given by $\chi_x = \prod_i a_i^{1-x_i} b_i^{x_i}.$ The final state of the quantum register has the form $ \sum_x \chi_x \theta_x |f(x)\rangle,$ where the $\theta_x$ are complex phases (which are efficiently computable as a function of $x$, but this will  not be relevant here) and where $f:\{0, 1\}^N \to \{0, 1\}^N$ is an invertible, efficiently computable Boolean function. Further, suppose that a subset $S$ of the $N$ qubits is measured, yielding as an outcome a bit string of $\mathbf{\alpha}=(\alpha_j: j\in S)$, where the bit $\alpha_{j}$ is the measurement outcome for qubit $j$. Then $\alpha$ occurs with probability $\pi(\alpha)=\sum |\chi_x|^2,$ where the sum is over all bit strings $x$ such that $f(x)_j = \alpha_j$ for all $j\in S$. To state it differently, $\pi(\alpha)$ is the total probability that an $N$-bit string $x$ which is generated with probability $p(x) = |\chi_x|^2$, satisfies $f(x)_j = \alpha_j$ for all $j\in S$. Now, it is possible to efficiently sample from the probability distribution $\{p(x)\}$. As a consequence, the following procedure (i)-(ii)-(iii) allows to efficiently generate a bit string $\alpha$ with probability $\pi(\alpha)$ on a classical computer: (i) generate an $N$-bit string $x$ with probability $p(x)=|\chi_x|^2$ (note that $\{p(x)\}$ is a simple product distribution since $|\chi\rangle$ is a product state); (ii) compute $f(x)$; (iii) set $\alpha_j$ to be the $j$-th coefficient of $f(x)$, for every $j\in S$. This shows it is possible to efficiently sample from the probability distribution $\{\pi(\alpha)\}$. 

Note that the first round, consisting of a local unitary operation, can be generalized while keeping the entire circuit classically simulatable. To do so, consider a quantum circuit ${\cal U}$ such that the computation $|\mathbf{0}\rangle\to{\cal U}|\mathbf{0}\rangle$, followed by a computational basis measurement of \emph{all} $N$ qubits, can be efficiently simulated classically in the weak sense. That is, it is classically possible to efficiently sample once from the probability distribution  $p(x) = |\langle x|{\cal U}|0\rangle|^2$. Then such a circuit, followed by an arbitrary (uniform) quantum circuit consisting of classical and diagonal gates only, and followed by a computational basis measurement of an arbitrary subset of the qubits, can be simulated efficiently classically in the weak sense. This last observation e.g. implies that a ``matchgate circuit'' \cite{Va01, Di04, Br08, Jo08}, a circuit of ``bounded tree-width'' \cite{Ma05, Sh06, Jo06}, a circuit which generates ``bounded Schmidt-rank'' \cite{Vi03, Va06}, or a quantum Fourier transform \cite{Ah06, Yo07, Br07}  (which are all known to be strongly simulatable), followed by an arbitrary poly-sized Toffoli-Diagonal circuit, is still weakly simulatable.

\section{Conclusion}

We have  studied classical simulation of quantum computation, taking the Gottesman-Knill theorem as a starting point. We have showed that each Clifford circuit ${\cal C}$ can be reduced to a simple equivalent circuit ${\cal C}'$ which, when applied to the input $|\mathbf{0}\rangle$, provides the same output state. Using this reduction to the normal form, we have attempted to provide a better understanding in the somewhat peculiar features of the Gottesman-Knill theorem. We have argued that the normal form provides insight in why the large amount of entanglement which can be generated by Clifford circuits fails to provide any quantum computational speed-up.  Furthermore, the normal form shows how the simulation of Clifford operations fits within the standard embedding of probabilistic classical computation into the quantum circuit model, related to the notion of HT circuits.

The class of HT circuits considered here, while indeed simple, exhibits features which are quite different from many results regarding classical simulation of quantum computation which have been found so far (such as the Gottesman-Knill theorem itself, matchgates, circuits of small tree-width, etc). For example, while such circuits are classically simulatable in the weak sense (i.e. they can be simulated using classical sampling techniques), strong classical simulation (i.e. the problem of \emph{computing} the output probabilities of measurement outcomes with high accuracy) is $\#\mathbf{P}$-complete and hence intractable. This is in contrast with previous results, the vast majority of which considering circuits where strong simulation is efficiently possible. Also, HT circuits can generate unbounded amounts of entanglement (as can e.g. Clifford circuits) while several previous results  use techniques which allow classical simulation of certain quantum circuits only in cases where the entanglement generated by these circuits is ``bounded'' (see e.g. \cite{Vi03, Ma05, Va06}). Thus, the present results provide simple examples of quantum circuits which can generate vast amounts of entanglement, which nevertheless does not result in any quantum computational speed-up.

We further note that a very  moderate extension of HT circuits immediately leads to highly non-trivial quantum algorithms. For example, it is known that Shor's factoring algorithm, using the phase estimation approach \cite{Ki95} (see also \cite{Sh05}), can be implemented by an HT circuit \emph{supplemented with a final round of Hadamard gates}, and then followed by a measurement in the computational basis. This shows that a mere final round of local basis changes allows to go from an HT circuit, which is trivially classically simulatable, to Shor's factoring algorithm, which is believed to provide an exponential speed-up.

{\bf Acknowledgements.} I am very grateful to R. Jozsa, W. D\"ur, H. Briegel, A. Kay and I. Cirac for interesting discussions and suggestions on the manuscript. Work supported by the Excellence Cluster MAP.

\section*{Appendix A: Updating the triple $(A, l, q)$}
Here we show the following claim: {\bf Claim:} suppose that an $N$-qubit state $|\psi\rangle$ is given in terms of an expression of the form (\ref{expansion}) for some $(A, l, q)$, and suppose that a gate $U\in\{H, P, \mbox{ CNOT}\}$ is applied to $|\psi\rangle$, resulting in a state $|\psi'\rangle$. Then $|\psi'\rangle$ still has the form (\ref{expansion}) for some updated $(A, l, q)$, and that these updates can be performed efficiently.

Before proceeding with the proof, we note the  following: in (\ref{expansion}), the  functions $q$ and $l$ are defined on the space of $N$-bit strings $\mathbf{Z}_2^N$. If the affine space $A$ is given by $A=\{Ru+t: u\in\mathbf{Z}_2^m\}$, we might as well write \be\label{expansion2}|\psi\rangle \propto \sum_{u} (-1)^{{\bar q}(u)} i^{ {\bar l}(u)}|Ru+t\rangle,\ee where now $\bar q$ and $\bar l$ are quadratic, resp. linear functions defined on the space of $m$-bit strings $u$, such that $\bar q(u) = q(Ru+t)$ and $\bar l(u)=l(Ru+t)$. Note that it is computationally easy to determine $\bar q$ ($\bar l$) from $q$ ($l$) and vice versa. In the following we will be a bit sloppy and simply identify $q(u)\equiv q(Ru+t)$ and similar for $l$; it will be clear from the context what the notation means.

We will also need the following lemma:

\begin{lem}\label{lemma}
Let $q(x)$ and $l(x)$ be quadratic and linear functions, resp., on $\mathbf{Z}_2^K$, where $x=(x_1, \dots, x_K)$. Denote $\bar x = (x_2, \dots, x_K)$. Then there exist a quadratic function $q'$ and  two linear functions $l'$ and $l''$ on $\mathbf{Z}_2^{K-1}$, such that \be \sum_{x_1} (-1)^{q(x)} i^{l(x)} \propto (-1)^{q'(\bar x)} i^{l'(\bar x)}\delta_{l''(\bar x), 0},\ee where $\propto$ denotes equality up to a multiplicative constant independent of $\bar x$. Moreover, determining $q', l'$ and $l''$ from $q$ and $l$ can be performed efficiently.
\end{lem}
The proof of the lemma is straightforward and is omitted. We now proceed with the proof of the claim.

{\bf Proof of claim:} First, suppose that $|\psi\rangle$ is given by (\ref{expansion}) and that a PHASE gate is applied to, say, the first qubit. As $P$ maps $|a\rangle$ to $i^a|a\rangle$ ($a=0,1$) the resulting state $|\psi'\rangle$ is \be|\psi'\rangle \propto \sum_{x\in A} (-1)^{{q}(x)} i^{ {l}(x)}i^{x_1}|x\rangle,\ee where $x_1$ is the first component of $x$. Using the identity\footnote{Here, the exponent $a+b$ is computed modulo 2.} $i^ai^b = (-1)^{ab} i^{a+b}$, for every $a, b\in \mathbf{Z}_2$, then shows that $|\psi'\rangle$ is again of the form (\ref{expansion}).

Second, suppose that a CNOT gate is applied to, say, the first and second qubit of $|\psi\rangle$. Note that CNOT maps $|a, b\rangle$ to $|a, a+b\rangle$, for every $a, b=0,1$. That is, CNOT performs a linear transformation (over $\mathbf{Z}_2$) ``within the ket''. This immediately implies that the state (\ref{expansion2}) is mapped to \be|\psi'\rangle \propto \sum_{u\in\mathbf{Z}_2^m} (-1)^{{q}(u)} i^{ {l}(u)}|R'u+t'\rangle,\ee for some appropriate $R'$ and $t'$ which are easily determined by performing the CNOT. Thus, also in this case the form (\ref{expansion}) is kept.

Finally, suppose that a Hadamard gate is applied to, say, the first qubit of $|\psi\rangle$. This is the most nontrivial case of the three. Recall that $H$ maps $|a\rangle$ to $\sum_{b=0}^1 (-1)^{ab}|b\rangle$ (where $a=0,1$). Denoting the first row of $R$ by $r^T$, letting $\bar R$ be the $(N-1)\times m$ matrix obtained by removing the first row of $R$ 
and denoting $\bar t=(t_2, \dots, t_N)$, one has \be\label{expansion_H} |\psi'\rangle \propto \sum_{v=0}^1\sum_{u\in\mathbf{Z}_2^m} (-1)^{q(u) + v\cdot (r^Tu) + vt_1}\ i^{l(u)} |v, \bar R u + \bar t\rangle.\ee Note that, as $R$ has full rank $m$, $\bar R$ may have either full rank $m$, or rank $m-1$. If $\bar R$ has full rank, then (\ref{expansion_H}) is of the form (\ref{expansion2}) and we are done. If $\bar R$ has rank $m-1$, some additional work is required. In this case, there is exactly one nontrivial linear combination of the columns of $\bar R$ which yields the identity. Without loss of generality, we may assume that the first column of $\bar R$ can be written as a linear combination of the other columns, and that the last $m-1$ columns of $\bar R$ are linearly independent. Denoting the columns of $\bar R$ by $c^i$ $(i=1, \dots, m)$, we therefore conclude that there exists a bit string $y=(y_2, \dots, y_{m})$ such that $c^1 = \sum_{i=2}^{m} y_i c^i$. Note that determining $y$ is a $\oplus\mathbf{L}$ problem as $y$ is the solution to a system of linear equations over $\mathbf{Z}_2$. It is then easy to verify that $\bar RQ = [0|c^2| \dots |c^m]$, where the $m\times m$ invertible matrix $Q$ is defined by \be Q = \left[\begin{array}{cccc} 1& & & \\y_2& 1& & \\\vdots & & \ddots& \\y_m & && 1\end{array}\right].\ee Making the substitution $u = Qu'$ in (\ref{expansion_H}) yields an expression of the following form: \be\label{expansion_H'} |\psi'\rangle \propto \sum_{v=0}^1\sum_{u'\in\mathbf{Z}_2^m} (-1)^{q'(v, u')}\ i^{l'(u')} |v, \bar RQu'  + \bar t\rangle,\ee for some quadratic and linear functions $q'$ and $l'$ which can be easily determined (by a $\oplus\mathbf{L}$ computer).  To lighten notation we will drop all primes in the above expression, i.e., $u'\equiv u$, $q'\equiv q$, $l'\equiv l$. Now note that the vector $\bar RQu$ does not depend on the first coefficient of $u$, since the first column of $\bar RQ$ is zero. Indeed, denoting $\bar u = (u_2, \dots, u_m)$, we have $\bar RQu = [c^2|\dots| c^m]\bar u$ (note also that $[c^2|\dots| c^m]$ has full rank). Therefore, in (\ref{expansion_H'}) the variable $u_1$ can be fully summed out. Using lemma \ref{lemma} then shows that $|\psi'\rangle$ is again of the desired form (\ref{expansion}). This completes the proof.

\end{document}